\documentclass[pra,aps,10pt,a4paper,nofootinbib,notitlepage,twocolumn,superscriptaddress,longbibliography]{revtex4-1}
\usepackage[utf8]{inputenc}
\usepackage[T1]{fontenc}
\usepackage[sc,osf]{mathpazo}
\usepackage[scaled=0.86]{berasans}
\usepackage[scaled=1.03]{inconsolata}
\usepackage[UKenglish]{babel}
\usepackage[unicode,colorlinks]{hyperref}
\usepackage{amsmath,amssymb,amsthm}
\usepackage{xspace}
\usepackage{mathtools}
\usepackage{pgfplots}
\pgfplotsset{compat=1.13}

\hypersetup{pdftitle={{Comment on "The photon identification loophole in EPRB experiments: computer models with single-wing selection"}}}

\begin{document}
\title{Comment on \\``The photon identification loophole in EPRB experiments: \\computer models with single-wing selection''}
\author{Mateus Araújo}
\affiliation{Institute for Theoretical Physics, University of Cologne, Zülpicher Straße 77, 50937 Cologne, Germany}
\author{Philippe Grangier}
\affiliation{Laboratoire Charles Fabry, IOGS, CNRS, Université Paris Saclay, F91127 Palaiseau, France}
\author{Jan-Åke Larsson}
\affiliation{Institutionen för systemteknik, Linköpings Universitet, 581 83 Linköping, Sweden}
\date{\today}

\begin{abstract}
 Contrary to the claims made in Ref.~\cite{deraedt17}, the recent Bell tests by Giustina et al.~\cite{giustina15} and Shalm et al.~\cite{shalm15} do not suffer from a ``photon identification loophole''. The model discussed in Ref.~\cite{deraedt17} is exploiting the well-known detection loophole, that is precisely what was closed in the recent experiments.
\end{abstract}

\maketitle

In Ref.~\cite{deraedt17}, the authors claim that due to the procedure used to identify which events correspond to a photon detection, the recent loophole-free Bell tests by Giustina et al.~\cite{giustina15} and Shalm et al.~\cite{shalm15} suffer from a ``photon identification loophole'' and are therefore inconclusive. To support this claim, the authors propose a hidden-variable model that is supposedly able to exploit this loophole in order to violate the Eberhard inequality  $J_\text{Eberhard} \geq 0$, where
\begin{equation}
\begin{split}\label{eq:eberhard}
 J_\text{Eberhard} = n_{oe}(\alpha_1,\beta_2) + n_{ou}(\alpha_1,\beta_2) + n_{eo}(\alpha_2,\beta_1) &\\ + n_{uo}(\alpha_2,\beta_1) + n_{oo}(\alpha_2,\beta_2) - n_{oo}(\alpha_1,\beta_1)&.
\end{split} 
\end{equation}
and all quantities are defined in Ref.~\cite{eberhard93}. 
The local hidden-variable model consists of sampling a real number $\lambda$ uniformly from the interval $[0,2\pi]$, and setting the local voltages of Alice and Bob's photons as
\begin{equation}
\begin{split}
 v_A &= r\sin^4\big(2(\alpha_x-\lambda)\big)/2 -1,\\
 v_B &= r'\sin^4\big(2(\beta_y-\lambda-\pi/2)\big)/2 -1,
\end{split} 
\end{equation}
where $\alpha_x,\beta_y$ are local detector settings, and $r,r'$ are random numbers sampled uniformly from the interval $[0,1]$. Each photon is considered detected if the local voltage is smaller than a threshold $V$, in which case the outcomes $a$ and $b$ of Alice and Bob are given by
\begin{equation}
\begin{split}
 a &= \operatorname{sign}[1+\cos\big(2(\alpha_x-\lambda)\big)-2r''], \\
 b &= \operatorname{sign}[1+\cos\big(2(\beta_y-\lambda-\pi/2)\big)-2r'''],
\end{split} 
\end{equation}
where $r'',r'''$ are again random numbers from the interval $[0,1]$. These outcomes are labelled $o=+1,e=-1$ for Alice and $o=-1,e=+1$ for Bob. If the photon is not detected, the outcome is labelled $u$. 

Each photon is detected with probability 
\begin{equation}
\begin{split}
 \eta_A(x,\lambda) &= \min\bigg\{1,\frac{2(V+1)}{\sin^4\big(2(\alpha_x-\lambda)\big)}\bigg\}, \\
 \eta_B(y,\lambda) &= \min\bigg\{1,\frac{2(V+1)}{\sin^4\big(2(\beta_y-\lambda-\pi/2)\big)}\bigg\},
\end{split} 
\end{equation}
that depends only on the local settings and the hidden variable $\lambda$. Therefore this threshold model is not exploiting a new loophole, but simply the well-known detection loophole. As such, it can violate inequalities that do not properly account for undetected photons, but cannot violate inequalities that do so. For instance, the Eberhard inequality \eqref{eq:eberhard} accounts for undetected photons by including the ``single'' counts $n_{ou}(\alpha_1,\beta_2)$ and $n_{uo}(\alpha_2,\beta_1)$, where one photon is counted but not the other.
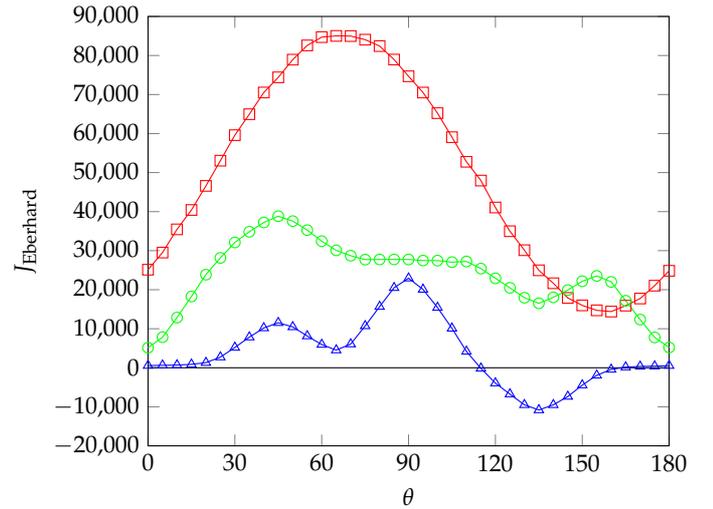
\begin{figure}[htpb]
	\begin{tikzpicture}
	\begin{axis}[
	xlabel=$\theta$,
	ylabel=$J_\text{Eberhard}$,
	xtick={0,30,60,90,120,150,180},
	ytick={-20000,-10000,0,10000,20000,30000,40000,50000,60000,70000,80000,90000},
	yticklabel style={/pgf/number format/fixed,/pgf/number format/precision=5},
	scaled y ticks=false,
	ymin=-20000, ymax=90000,
	xmin=0,xmax=180,
	legend style={at={(0.7,0.8)},
	anchor=north west},
	]

	\addplot[mark=none, black] coordinates {(0,0) (180,0)};
	
	\addplot[color=red,mark=square] file {eberhard_without_threshold};

	\addplot[color=green,mark=o] file {eberhard_with_threshold};

	\addplot[color=blue,mark=triangle] file {deleted_with_threshold};

	\end{axis}
	\end{tikzpicture}
\caption{Results of the simulation without thresholding (red squares), with thresholding (green circles), and with thresholding and deleting the terms $n_{ou}(\alpha_1,\beta_2)$ and $n_{uo}(\alpha_2,\beta_1)$ (blue triangles). The Eberhard inequality $J_\text{Eberhard} \geq 0$ is obeyed in the two first cases, whereas in the third case the violation of the modified inequality is due to the detection loophole. }
\label{fig:eberhard}
\end{figure}

The authors claim, nevertheless, that their model can violate the Eberhard inequality, and present numerical results purporting to show this. As a double check, we implemented their hidden-variable model (our code is available as an ancillary file), and ran the program with the same settings: $\alpha_1 = \theta + 3\pi/8$, $\alpha_2 = \theta + \pi/8$, $\beta_1 = \pi/8$, and $\beta_2 = 3\pi/8$. The results are shown in Fig.~\ref{fig:eberhard}: the red squares show the case of threshold $V=0$, implying perfect efficiency, and the green circles show the case of threshold $V=-0.995$  as used in Ref.~\cite{deraedt17}, implying imperfect efficiency. We calculated the average efficiency to be $\bar\eta = (6+50\arcsin(1/\sqrt{10}))/25\pi \approx 0.28$. Each point consists of $4\times 10^5$ trials.

In order to reproduce the claimed violation, represented here by the blue triangles, we had to set $n_{ou}(\alpha_1,\beta_2) = n_{uo}(\alpha_2,\beta_1) = 0$, whereas the model gives $n_{ou}(\alpha_1,\beta_2) > 0$ and $n_{uo}(\alpha_2,\beta_1) > 0$.  The exact procedure for producing the graph of Ref.~\cite{deraedt17} cannot be found in the paper, but the authors told us in a private communication that they also took $n_{ou}(\alpha_1,\beta_2) = n_{uo}(\alpha_2,\beta_1) = 0$, so all the simulation data is fully consistent.  We will not discuss here the reasons for their decision to delete counts where one photon remains undetected ($u$). We only point out that the deletion of these terms changes the tested inequality from Eberhard's into a different inequality that is vulnerable to the detection loophole. 

As a conclusion, and as it can be seen in Fig.~\ref{fig:eberhard}, when all relevant counts are taken into account there is clearly no violation of the Eberhard inequality via the thresholding mechanism -- as it must be. 

\bibliography{biblio}

\begin{thebibliography}{4}%
\makeatletter
\providecommand \@ifxundefined [1]{%
 \@ifx{#1\undefined}
}%
\providecommand \@ifnum [1]{%
 \ifnum #1\expandafter \@firstoftwo
 \else \expandafter \@secondoftwo
 \fi
}%
\providecommand \@ifx [1]{%
 \ifx #1\expandafter \@firstoftwo
 \else \expandafter \@secondoftwo
 \fi
}%
\providecommand \natexlab [1]{#1}%
\providecommand \enquote  [1]{``#1''}%
\providecommand \bibnamefont  [1]{#1}%
\providecommand \bibfnamefont [1]{#1}%
\providecommand \citenamefont [1]{#1}%
\providecommand \href@noop [0]{\@secondoftwo}%
\providecommand \href [0]{\begingroup \@sanitize@url \@href}%
\providecommand \@href[1]{\@@startlink{#1}\@@href}%
\providecommand \@@href[1]{\endgroup#1\@@endlink}%
\providecommand \@sanitize@url [0]{\catcode `\\12\catcode `\$12\catcode
  `\&12\catcode `\#12\catcode `\^12\catcode `\_12\catcode `\%12\relax}%
\providecommand \@@startlink[1]{}%
\providecommand \@@endlink[0]{}%
\providecommand \url  [0]{\begingroup\@sanitize@url \@url }%
\providecommand \@url [1]{\endgroup\@href {#1}{\urlprefix }}%
\providecommand \urlprefix  [0]{URL }%
\providecommand \Eprint [0]{\href }%
\providecommand \doibase [0]{http://dx.doi.org/}%
\providecommand \selectlanguage [0]{\@gobble}%
\providecommand \bibinfo  [0]{\@secondoftwo}%
\providecommand \bibfield  [0]{\@secondoftwo}%
\providecommand \translation [1]{[#1]}%
\providecommand \BibitemOpen [0]{}%
\providecommand \bibitemStop [0]{}%
\providecommand \bibitemNoStop [0]{.\EOS\space}%
\providecommand \EOS [0]{\spacefactor3000\relax}%
\providecommand \BibitemShut  [1]{\csname bibitem#1\endcsname}%
\let\auto@bib@innerbib\@empty
\bibitem [{\citenamefont {{De Raedt}}\ \emph {et~al.}(2017)\citenamefont {{De
  Raedt}}, \citenamefont {{Michielsen}},\ and\ \citenamefont
  {{Hess}}}]{deraedt17}%
  \BibitemOpen
  \bibfield  {author} {\bibinfo {author} {\bibfnamefont {H.}~\bibnamefont {{De
  Raedt}}}, \bibinfo {author} {\bibfnamefont {K.}~\bibnamefont {{Michielsen}}},
  \ and\ \bibinfo {author} {\bibfnamefont {K.}~\bibnamefont {{Hess}}},\
  }\bibfield  {title} {\enquote {\bibinfo {title} {{The photon identification
  loophole in EPRB experiments: computer models with single-wing selection}},}\
  }\href {\doibase 10.1515/phys-2017-0085} {\bibfield  {journal} {\bibinfo
  {journal} {Open Physics}\ }\textbf {\bibinfo {volume} {15}},\ \bibinfo {eid}
  {85} (\bibinfo {year} {2017})},\ \Eprint {http://arxiv.org/abs/1707.08307}
  {arXiv:1707.08307 [quant-ph]} \BibitemShut {NoStop}%
\bibitem [{\citenamefont {{Giustina}}\ \emph {et~al.}(2015)\citenamefont
  {{Giustina}}, \citenamefont {{Versteegh}}, \citenamefont {{Wengerowsky}},
  \citenamefont {{Handsteiner}}, \citenamefont {{Hochrainer}}, \citenamefont
  {{Phelan}}, \citenamefont {{Steinlechner}}, \citenamefont {{Kofler}},
  \citenamefont {{Larsson}}, \citenamefont {{Abell{\'a}n}}, \citenamefont
  {{Amaya}}, \citenamefont {{Pruneri}}, \citenamefont {{Mitchell}},
  \citenamefont {{Beyer}}, \citenamefont {{Gerrits}}, \citenamefont {{Lita}},
  \citenamefont {{Shalm}}, \citenamefont {{Nam}}, \citenamefont {{Scheidl}},
  \citenamefont {{Ursin}}, \citenamefont {{Wittmann}},\ and\ \citenamefont
  {{Zeilinger}}}]{giustina15}%
  \BibitemOpen
  \bibfield  {author} {\bibinfo {author} {\bibfnamefont {M.}~\bibnamefont
  {{Giustina}}}, \bibinfo {author} {\bibfnamefont {M.~A.~M.}\ \bibnamefont
  {{Versteegh}}}, \bibinfo {author} {\bibfnamefont {S.}~\bibnamefont
  {{Wengerowsky}}}, \bibinfo {author} {\bibfnamefont {J.}~\bibnamefont
  {{Handsteiner}}}, \bibinfo {author} {\bibfnamefont {A.}~\bibnamefont
  {{Hochrainer}}}, \bibinfo {author} {\bibfnamefont {K.}~\bibnamefont
  {{Phelan}}}, \bibinfo {author} {\bibfnamefont {F.}~\bibnamefont
  {{Steinlechner}}}, \bibinfo {author} {\bibfnamefont {J.}~\bibnamefont
  {{Kofler}}}, \bibinfo {author} {\bibfnamefont {J.-{\AA}.}\ \bibnamefont
  {{Larsson}}}, \bibinfo {author} {\bibfnamefont {C.}~\bibnamefont
  {{Abell{\'a}n}}}, \bibinfo {author} {\bibfnamefont {W.}~\bibnamefont
  {{Amaya}}}, \bibinfo {author} {\bibfnamefont {V.}~\bibnamefont {{Pruneri}}},
  \bibinfo {author} {\bibfnamefont {M.~W.}\ \bibnamefont {{Mitchell}}},
  \bibinfo {author} {\bibfnamefont {J.}~\bibnamefont {{Beyer}}}, \bibinfo
  {author} {\bibfnamefont {T.}~\bibnamefont {{Gerrits}}}, \bibinfo {author}
  {\bibfnamefont {A.~E.}\ \bibnamefont {{Lita}}}, \bibinfo {author}
  {\bibfnamefont {L.~K.}\ \bibnamefont {{Shalm}}}, \bibinfo {author}
  {\bibfnamefont {S.~W.}\ \bibnamefont {{Nam}}}, \bibinfo {author}
  {\bibfnamefont {T.}~\bibnamefont {{Scheidl}}}, \bibinfo {author}
  {\bibfnamefont {R.}~\bibnamefont {{Ursin}}}, \bibinfo {author} {\bibfnamefont
  {B.}~\bibnamefont {{Wittmann}}}, \ and\ \bibinfo {author} {\bibfnamefont
  {A.}~\bibnamefont {{Zeilinger}}},\ }\bibfield  {title} {\enquote {\bibinfo
  {title} {{Significant-Loophole-Free Test of Bell's Theorem with Entangled
  Photons}},}\ }\href {\doibase 10.1103/PhysRevLett.115.250401} {\bibfield
  {journal} {\bibinfo  {journal} {Phys. Rev. Lett.}\ }\textbf {\bibinfo
  {volume} {115}},\ \bibinfo {pages} {250401} (\bibinfo {year} {2015})},\
  \Eprint {http://arxiv.org/abs/1511.03190} {arXiv:1511.03190 [quant-ph]}
  \BibitemShut {NoStop}%
\bibitem [{\citenamefont {{Shalm}}\ \emph {et~al.}(2015)\citenamefont
  {{Shalm}}, \citenamefont {{Meyer-Scott}}, \citenamefont {{Christensen}},
  \citenamefont {{Bierhorst}}, \citenamefont {{Wayne}}, \citenamefont
  {{Stevens}}, \citenamefont {{Gerrits}}, \citenamefont {{Glancy}},
  \citenamefont {{Hamel}}, \citenamefont {{Allman}}, \citenamefont {{Coakley}},
  \citenamefont {{Dyer}}, \citenamefont {{Hodge}}, \citenamefont {{Lita}},
  \citenamefont {{Verma}}, \citenamefont {{Lambrocco}}, \citenamefont
  {{Tortorici}}, \citenamefont {{Migdall}}, \citenamefont {{Zhang}},
  \citenamefont {{Kumor}}, \citenamefont {{Farr}}, \citenamefont {{Marsili}},
  \citenamefont {{Shaw}}, \citenamefont {{Stern}}, \citenamefont
  {{Abell{\'a}n}}, \citenamefont {{Amaya}}, \citenamefont {{Pruneri}},
  \citenamefont {{Jennewein}}, \citenamefont {{Mitchell}}, \citenamefont
  {{Kwiat}}, \citenamefont {{Bienfang}}, \citenamefont {{Mirin}}, \citenamefont
  {{Knill}},\ and\ \citenamefont {{Nam}}}]{shalm15}%
  \BibitemOpen
  \bibfield  {author} {\bibinfo {author} {\bibfnamefont {L.~K.}\ \bibnamefont
  {{Shalm}}}, \bibinfo {author} {\bibfnamefont {E.}~\bibnamefont
  {{Meyer-Scott}}}, \bibinfo {author} {\bibfnamefont {B.~G.}\ \bibnamefont
  {{Christensen}}}, \bibinfo {author} {\bibfnamefont {P.}~\bibnamefont
  {{Bierhorst}}}, \bibinfo {author} {\bibfnamefont {M.~A.}\ \bibnamefont
  {{Wayne}}}, \bibinfo {author} {\bibfnamefont {M.~J.}\ \bibnamefont
  {{Stevens}}}, \bibinfo {author} {\bibfnamefont {T.}~\bibnamefont
  {{Gerrits}}}, \bibinfo {author} {\bibfnamefont {S.}~\bibnamefont {{Glancy}}},
  \bibinfo {author} {\bibfnamefont {D.~R.}\ \bibnamefont {{Hamel}}}, \bibinfo
  {author} {\bibfnamefont {M.~S.}\ \bibnamefont {{Allman}}}, \bibinfo {author}
  {\bibfnamefont {K.~J.}\ \bibnamefont {{Coakley}}}, \bibinfo {author}
  {\bibfnamefont {S.~D.}\ \bibnamefont {{Dyer}}}, \bibinfo {author}
  {\bibfnamefont {C.}~\bibnamefont {{Hodge}}}, \bibinfo {author} {\bibfnamefont
  {A.~E.}\ \bibnamefont {{Lita}}}, \bibinfo {author} {\bibfnamefont {V.~B.}\
  \bibnamefont {{Verma}}}, \bibinfo {author} {\bibfnamefont {C.}~\bibnamefont
  {{Lambrocco}}}, \bibinfo {author} {\bibfnamefont {E.}~\bibnamefont
  {{Tortorici}}}, \bibinfo {author} {\bibfnamefont {A.~L.}\ \bibnamefont
  {{Migdall}}}, \bibinfo {author} {\bibfnamefont {Y.}~\bibnamefont {{Zhang}}},
  \bibinfo {author} {\bibfnamefont {D.~R.}\ \bibnamefont {{Kumor}}}, \bibinfo
  {author} {\bibfnamefont {W.~H.}\ \bibnamefont {{Farr}}}, \bibinfo {author}
  {\bibfnamefont {F.}~\bibnamefont {{Marsili}}}, \bibinfo {author}
  {\bibfnamefont {M.~D.}\ \bibnamefont {{Shaw}}}, \bibinfo {author}
  {\bibfnamefont {J.~A.}\ \bibnamefont {{Stern}}}, \bibinfo {author}
  {\bibfnamefont {C.}~\bibnamefont {{Abell{\'a}n}}}, \bibinfo {author}
  {\bibfnamefont {W.}~\bibnamefont {{Amaya}}}, \bibinfo {author} {\bibfnamefont
  {V.}~\bibnamefont {{Pruneri}}}, \bibinfo {author} {\bibfnamefont
  {T.}~\bibnamefont {{Jennewein}}}, \bibinfo {author} {\bibfnamefont {M.~W.}\
  \bibnamefont {{Mitchell}}}, \bibinfo {author} {\bibfnamefont {P.~G.}\
  \bibnamefont {{Kwiat}}}, \bibinfo {author} {\bibfnamefont {J.~C.}\
  \bibnamefont {{Bienfang}}}, \bibinfo {author} {\bibfnamefont {R.~P.}\
  \bibnamefont {{Mirin}}}, \bibinfo {author} {\bibfnamefont {E.}~\bibnamefont
  {{Knill}}}, \ and\ \bibinfo {author} {\bibfnamefont {S.~W.}\ \bibnamefont
  {{Nam}}},\ }\bibfield  {title} {\enquote {\bibinfo {title} {{Strong
  Loophole-Free Test of Local Realism$^{*}$}},}\ }\href {\doibase
  10.1103/PhysRevLett.115.250402} {\bibfield  {journal} {\bibinfo  {journal}
  {Phys. Rev. Lett.}\ }\textbf {\bibinfo {volume} {115}},\ \bibinfo {pages}
  {250402} (\bibinfo {year} {2015})},\ \Eprint
  {http://arxiv.org/abs/1511.03189} {arXiv:1511.03189 [quant-ph]} \BibitemShut
  {NoStop}%
\bibitem [{\citenamefont {Eberhard}(1993)}]{eberhard93}%
  \BibitemOpen
  \bibfield  {author} {\bibinfo {author} {\bibfnamefont {P.~H.}\ \bibnamefont
  {Eberhard}},\ }\bibfield  {title} {\enquote {\bibinfo {title} {Background
  level and counter efficiencies required for a loophole-free
  einstein-podolsky-rosen experiment},}\ }\href {\doibase
  10.1103/PhysRevA.47.R747} {\bibfield  {journal} {\bibinfo  {journal} {Phys.
  Rev. A}\ }\textbf {\bibinfo {volume} {47}},\ \bibinfo {pages} {R747--R750}
  (\bibinfo {year} {1993})}\BibitemShut {NoStop}%
\end{thebibliography}%

\end{document}